%% file: paper.tex
\documentclass[runningheads,a4paper,oribibl]{llncs}[2015/06/24]

\pdfoutput=1

\usepackage{cmap}
\usepackage{doi}

\iftrue 
  \usepackage[%
    rm={oldstyle=false,proportional=true},%
    sf={oldstyle=false,proportional=true},%
    tt={oldstyle=false,proportional=true,variable=true},%
    qt=false%
  ]{cfr-lm}
\else
  \usepackage{newtxtext}
  \usepackage{newtxmath}
  \usepackage[zerostyle=b,scaled=.9]{newtxtt}
\fi

\usepackage[math]{blindtext}
\usepackage{mwe}

\usepackage{pdflscape}

\usepackage{amsthm}
\theoremstyle{remark}
\usepackage{mdframed}
\usepackage{lipsum}
\newmdtheoremenv{thesis}{Thesis}

\usepackage[T1]{fontenc}
\usepackage[utf8]{inputenc} 

\usepackage{graphicx}

\usepackage[ngerman,english]{babel}
\addto\extrasenglish{\languageshorthands{ngerman}\useshorthands{"}}

\usepackage{upquote}

\usepackage{booktabs}

\usepackage{paralist}

\usepackage{csquotes}

\RequirePackage{iftex}
\ifPDFTeX
  \RequirePackage[%
    final,%
    expansion=alltext,%
    protrusion=alltext-nott]{microtype}%
\else
  \RequirePackage[%
    final,%
    protrusion=alltext-nott]{microtype}%
\fi%
\DisableLigatures{encoding = T1, family = tt* }

\usepackage{url}
\makeatletter
\g@addto@macro{\UrlBreaks}{\UrlOrds}
\makeatother

\usepackage{xcolor}

\usepackage{listings}
\renewcommand{\lstlistingname}{List.}
\usepackage{chngcntr}
\AtBeginDocument{\counterwithout{lstlisting}{section}}

\AtBeginEnvironment{listing}{\setcounter{listing}{\value{lstlisting}}}
\AtEndEnvironment{listing}{\stepcounter{lstlisting}}

\usepackage{pdfcomment}
\usepackage[%
  square,        
  comma,         
  numbers,       
  sort&compress, 
]{natbib}

\SetExpansion
[ context = sloppy,
  stretch = 30,
  shrink = 60,
  step = 5 ]
{ encoding = {OT1,T1,TS1} }
{ }

\usepackage{hyperref}
\hypersetup{hidelinks,
  colorlinks=true,
  allcolors=black,
  pdfstartview=Fit,
  breaklinks=true}
\usepackage[all]{hypcap}

\usepackage[capitalise,nameinlink]{cleveref}
\crefname{section}{Sect.}{Sect.}
\Crefname{section}{Section}{Sections}
\crefname{listing}{\lstlistingname}{\lstlistingname}
\Crefname{listing}{Listing}{Listings}

\usepackage{ltxcmds}
\makeatletter
\newcommand{\IfPackageLoaded}[2]{\ltx@ifpackageloaded{#1}{#2}{}}
\makeatother

\IfPackageLoaded{minted}{
  \setminted{numbersep=5pt, xleftmargin=12pt}

  \usemintedstyle{friendly}

  \usepackage[labelfont=bf,font=small,skip=4pt]{caption}
  \SetupFloatingEnvironment{listing}{name=List.,within=none}
}{
}
\ifcsmacro{listing}{}{
  \newenvironment{listing}[1][htbp!]{\begin{figure}[#1]}{\end{figure}}
  \newcounter{listing}
}
\ifcsmacro{minted}{}{%
  
}

\DeclareFontFamily{U}{MnSymbolC}{}
\DeclareSymbolFont{MnSyC}{U}{MnSymbolC}{m}{n}
\DeclareFontShape{U}{MnSymbolC}{m}{n}{
  <-6>    MnSymbolC5
  <6-7>   MnSymbolC6
  <7-8>   MnSymbolC7
  <8-9>   MnSymbolC8
  <9-10>  MnSymbolC9
  <10-12> MnSymbolC10
  <12->   MnSymbolC12%
}{}
\DeclareMathSymbol{\powerset}{\mathord}{MnSyC}{180}

\begin{document}

\title{Future Directions for Optimizing Compilers%
}


\author{Nuno P. Lopes\inst{1}\and John Regehr\inst{2}}

\institute{%
  Microsoft Research, UK\\
  \email{nlopes@microsoft.com}\and
  University of Utah, USA\\
  \email{regehr@cs.utah.edu}}

%
%

\maketitle

\noindent\makebox[\linewidth]{{September 6, 2018}}

\input{intro}
\input{mainidea}
\input{declarative}
\input{souper}
\input{moresuperoptimizers}
\input{data}
\input{semantics}
\input{further}
\input{ir}
\input{conclusions}

\subsection*{Acknowledgments}

The authors would like to thank Dan Gohman and members of the Utah PL
reading group for their valuable feedback on drafts of this paper.

\bibliographystyle{plain}
\begingroup
  \microtypecontext{expansion=sloppy}
  \bibliography{paper}
\endgroup
\end{document}

%% file: intro.tex
\section{Introduction}

As software becomes larger, programming languages become higher-level,
and processors continue to fail to be clocked faster, we'll
increasingly require compilers to reduce code bloat, eliminate
abstraction penalties, and exploit interesting instruction sets.
At the same time, compiler execution time must not increase too much
and also compilers should never produce the wrong output.
This paper examines the problem of making optimizing compilers faster,
less buggy, and more capable of generating high-quality output.

\subsection{Why are Compilers Slow?}

While very fast compilers
exist,\footnote{\url{https://bellard.org/tcc/tccboot.html}} heavily
optimizing ahead-of-time compilers are generally not fast.
First, many of the sub-problems that compilers are trying to solve,
such as optimal instruction selection, are themselves intractable.
Second, after performing basic optimizations that are always a good
idea, and that usually reduce code size, a compiler is confronted with
optimization opportunities (auto-vectorization, loop unrolling,
function inlining, etc.) that have a less clear payoff and that often
increase the amount of code that has to be subsequently processed.
Third, the compiler would like to reach a fixed point where every
obviously desirable transformation has been performed; there is rarely
enough time to do this.
Finally, fully breaking down the abstractions in high-level-language
code is not an easy job.

\subsection{Why are Compilers Wrong?}
\label{sec:wrong}

Compiler bugs have diverse root causes:
\begin{itemize}
\item
Difficult-to-avoid pitfalls in unsafe compiler implementation
languages---commonly, null pointer dereferences and use-after-free
errors.
\item
Little-used and unclear corner cases in the language standards.
For example, C and C++'s integer promotion
rules\footnote{\url{https://blog.regehr.org/archives/482}} and
volatile qualifier~\cite{Eide08} have caused trouble.
\item
Little-used and unclear corner cases in the compiler intermediate
representation (IR) semantics.
To support aggressive optimizations, compiler IRs have evolved
sophisticated notions about undefined behavior that are error prone
and also have typically not been formally specified or even documented
adequately~\cite{Lee17}.
\item
Failure to correctly handle corner cases while implementing an
optimization.
Peephole optimizations, in particular, seem difficult for people to
reason about; LLVM's peephole optimizer was its buggiest file
according to Csmith~\cite{Yang11}.
\item
Failure to respect an invariant on a compiler data structure.
These invariants are often quite sophisticated and are not always
well-documented.
\item
Complexity and shortcuts that are a consequence of trying to make the
compiler go fast.
For example, caching results instead of recomputing them is
error-prone when the results may be invalidated in a fine-grained way.
\item
Code churn due to external requirements such as new language
standards, new targets and target features, and improved
optimizations.
The global impact of new externally-motivated features is not always
clear at first.
\end{itemize}
Taken together, these factors seem to make compiler bugs a fact of
life.
Also, while compilers are on the face of it eminently testable (their
job is well-understood, they are deterministic, they run reasonably
quickly, and they have few external dependencies), they often seem to
be under-tested, since buggy behaviors can usually be triggered by
small, innocuous-looking inputs.

\subsection{Why is the Quality of Generated Code Sometimes Poor?}

An optimizing compiler is faced with a collection of intractable
search problems and given very little time in which to solve them.
The solutions to these problems enable and block each other in ways
that can be difficult to predict ahead of time: this is the ``phase
ordering'' problem where a compiler must find a sequence of
optimization passes that seems to give good results most of the time,
without taking too long to execute.
Optimizations that work well in one problem domain (loop vectorization
or aggressive partial evaluation, for example) are often useless or
counterproductive in other domains, and yet a given compiler instance
has no obvious principled way to know what will work well this time.

Moving up a level of abstraction, compiler development teams are faced
with many demands on their time---fixing bugs, supporting new
language features, supporting new target features, dealing with
Spectre and Meltdown, etc.---besides working on new optimizations.
Little has been written about the economics of compiler optimization
(Robison's 2001 paper~\cite{Robison01} is the exception), but it is
obvious that economics is decisive in determining what we can
accomplish in this space.

An important part of high-quality code generation is optimizing
predictably: a computation should be converted into high-quality
machine code regardless of its context and how it is expressed in the
language syntax.
A partial solution to predictable optimization is aggressive
normalization of IR\@.
Lawrence~\cite{Lawrence07} says:
\begin{quote}
Normalization refers to where many different source programs result in
the same machine code after compilation---that is, where the same
machine code is produced regardless of which form was written by the
programmer.
\end{quote}
Normalization is an important, useful property of an IR\@.
However, in this paper we explore a different approach to
predictability: finding optimizations using aggressive search
algorithms and SMT solvers that are, unlike humans, not easy to fool
with superficial changes of representation.

Proebsting's
Law\footnote{\url{http://proebsting.cs.arizona.edu/law.html}} states
that performance gains due to improvements in compiler optimizations
will double the speed of a program every 18 years.
This allusion to Moore's Law was funnier back when we were seeing
massive performance increases every year due to hardware.
Nevertheless, Proebsting's Law---which comes out to 4\% per
year---likely overstates performance gains due to compiler
optimizations.
The obvious experiment that would validate or refute this law is easy
to perform over a span of a few years, but would be difficult over
longer time frames since a fixed platform and benchmark suite has to
be supported by the compiler under test for the duration of the
experiment.
An obvious but wrong way to examine Proebsting's Law would be to look
at the size of the performance gap between unoptimized and optimized
code over time.
This doesn't work because ``unoptimized'' doesn't have an objective
meaning.

If even 4\% per year is too much to expect from compiler technology,
is it still worth working on optimizers?
Pugh\footnote{\url{http://www.cs.umd.edu/~pugh/IsCodeOptimizationRelevant.pdf}}
and
Bernstein\footnote{\url{http://cr.yp.to/talks/2015.04.16/slides-djb-20150416-a4.pdf}}
have argued that it isn't.
On the other hand, given users' evident preference for fast code, it
isn't clear that these arguments---which were perhaps unserious to
begin with---hold any water, and in any case getting high-performance
executables out of high-level languages seems to fundamentally require
aggressive compiler optimization.


%% file: mainidea.tex
\section{Semantics, Solvers, Synthesis, and Superoptimizers}

This paper is organized around a handful of thesis statements; the main
one is:
\begin{thesis}
Major components of future compilers will be generated
partially automatically, with the help of SMT solvers, directly
addressing compiler speed (automatically discovered optimizations will
be structured uniformly and amenable to fast rewrite strategies),
compiler correctness (automated theorem provers are generally less
fallible than humans), and code quality (solvers can be used to
conduct a thorough search of the optimization space).
\end{thesis}
Key ingredients in realizing this idea are formal semantics for the
languages being optimized, use of SMT solvers to verify equivalence or
refinement, and effective synthesis algorithms to create optimized
code in cases where brute-force search doesn't work.


An easy place to start is the collection of rewrite rules that most
optimizing compilers use to remove local inefficiencies.
For example, this line from the Go
compiler:\footnote{\url{https://github.com/golang/go/blob/go1.10.1/src/cmd/compile/internal/ssa/gen/generic.rules\#L661}}

{\small\begin{verbatim}
(Xor64 x (Xor64 x y)) -> y
\end{verbatim}}

looks for code like this:

{\small\begin{verbatim}
func foo(x int, y int) int {
    return x ^ (x ^ y)
}
\end{verbatim}}

and optimizes it to simply return \texttt{y}.
The optimization isn't written in Go, but rather in a domain-specific
language (DSL) for pattern rewrites.
At the time the Go compiler is compiled, the DSL is translated into Go
and then compiled along with the rest of the compiler source code.
The compiler for the Mesa shader language contains an analogous
collection of rules, such as this one that takes advantage of one of
De~Morgan's
Laws:\footnote{\url{https://github.com/mesa3d/mesa/blob/master/src/compiler/nir/nir_opt_algebraic.py\#L351}}

{\small\begin{verbatim}
(('iand', ('inot', a), ('inot', b)), ('inot', ('ior',  a, b)))
\end{verbatim}}

In other words, $\neg a \wedge \neg b$ can be rewritten as $\neg(a\vee
b)$, saving one operation.
GCC specifies similar transformations using a Lispish language:%
\footnote{\url{https://github.com/gcc-mirror/gcc/blob/gcc-8_1_0-release/gcc/match.pd\#L709}\\%
\url{https://gcc.gnu.org/onlinedocs/gccint/The-Language.html}}

{\small\begin{verbatim}
/* PR53979: Transform ((a ^ b) | a) -> (a | b) */
(simplify
  (bit_ior:c (bit_xor:c @0 @1) @0)
  (bit_ior @0 @1))
\end{verbatim}}

Alas, not all compilers have lifted their peephole optimizations into
a form that can be conveniently separated from the rest of the
compiler.
For example, LuaJIT has 2,300 lines of C to do this
job,\footnote{\url{https://github.com/LuaJIT/LuaJIT/blob/master/src/lj_opt_fold.c}}
libFirm has 8000
lines,\footnote{\url{https://github.com/libfirm/libfirm/blob/master/ir/opt/iropt.c}}
and LLVM's instruction combiner (InstCombine) probably wins the prize
for the strongest and most baroque IR-level peephole optimizer ever
created, at 30,000 lines of
C++.\footnote{\url{https://github.com/llvm-mirror/llvm/tree/master/lib/Transforms/InstCombine}}
Also, GCC has many transformations that aren't specified
in \texttt{match.pd} but rather appear explicitly in code.

Two threads of research will help us improve upon existing
peephole optimizers.
First, compiler developers and researchers should:
\begin{enumerate}
\item
Create a declarative language for writing IR-level optimizations.
\item
Implement solver-based tools for finding incorrect optimizations,
not-weakest preconditions, groups of optimizations that either subsume
each other or undo each other, etc.
\item
Implement a compiler-compiler with the goal of generating specialized
code that can perform the specified optimizations scalably: the
runtime of the optimizer should be a sublinear function of the number
of optimizations and the constant factor should be small. This is
important for both JIT and AOT compilers.
\end{enumerate}
Although several compilers (as we have seen) have taken step~1 in this
list, much work remains to be done for items~2 and~3.
A highly useful side effect of writing a verifier is that it forces a
formal specification of IR semantics to be written.
Most IRs are only informally specified, guaranteeing the existence of
dark corners in the semantics.

The second research thread is to:
\begin{enumerate}
\item
Derive new optimizations either from first principles or by looking
for optimizations that are missed in practice. The details of
the search procedure aren't important: it could be randomized,
enumeration-based, or counter-example guided inductive synthesis
(CEGIS).
The tool that discovers optimizations using an expensive search is
called a superoptimizer.
\item
Express the discovered optimizations in a suitably general form, with
appropriate preconditions, in the declarative language.
\item
Iterate until a large fraction of expressible, profitable
optimizations are performed by the compiler.
\item
Increase the expressiveness and reach of declarative optimizations,
with the goal of replacing more and more hand-written compiler code
over time.
Similar ideas can be used to help automate the construction of parts
of the compiler other than peephole optimizations, but we'll discuss
them later.
\end{enumerate}
Here, again, there has been some progress but much work remains.

\begin{figure}[tbp]
\begin{center}
\includegraphics[width=0.80\textwidth]{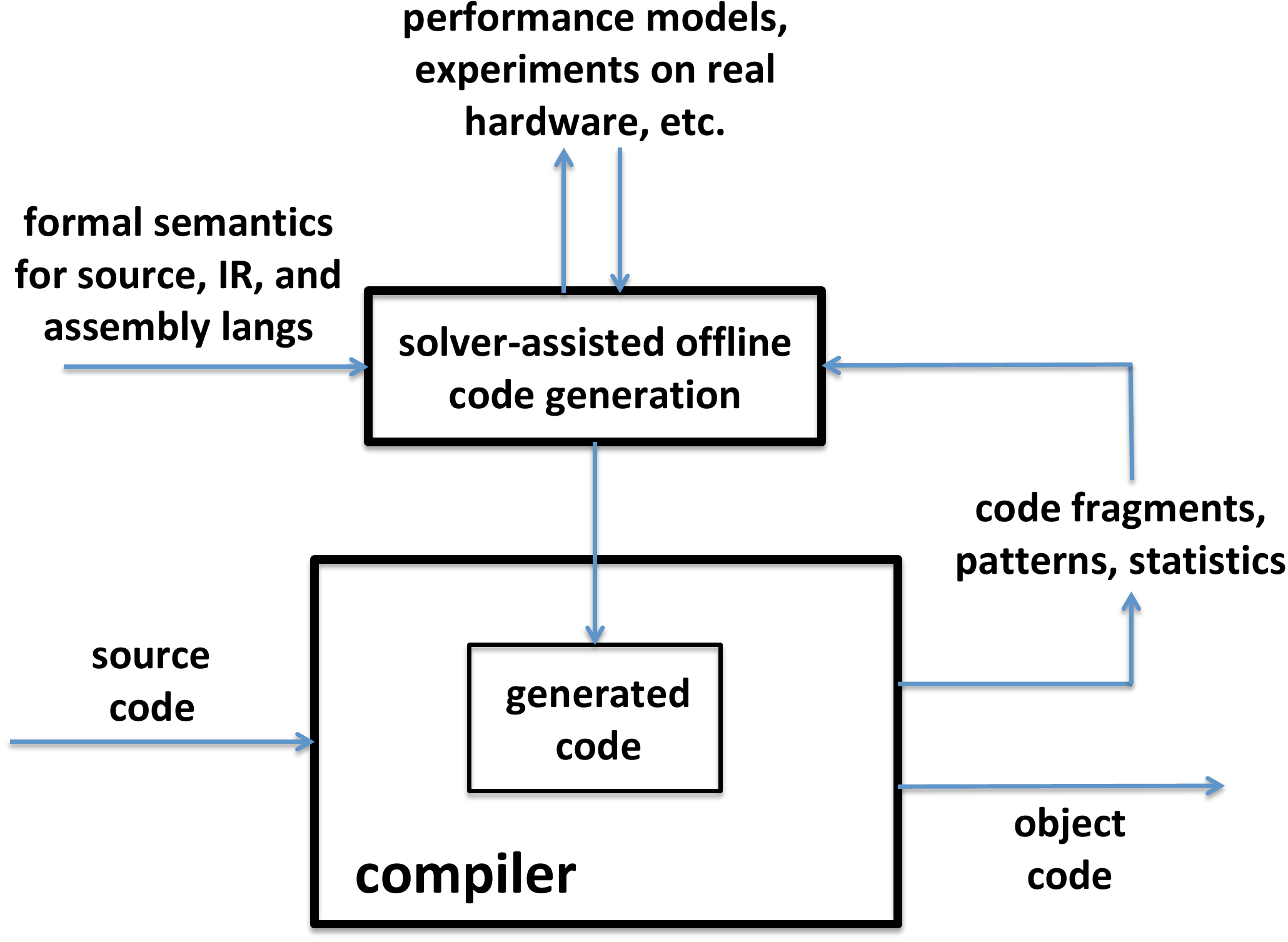}
\end{center}
\caption{The big picture}
\label{fig:big}
\end{figure}

Figure~\ref{fig:big} shows the big picture: a feedback loop that makes
the compiler more effective at compiling the code it is given.
The proposed feedback loop is unlike profile-guided optimization,
which optimizes a program based on its own observed execution
characteristics.
The proposed feedback loop is also dissimilar to most existing
research on applying machine learning to compilers, which typically
focuses on improving heuristics (phase ordering, function inlining
thresholds, etc.) rather than on deriving entirely new optimizations.

This overall vision is not new, but rather builds on a compact body of
work that perhaps began in 1979 when Fraser~\cite{Fraser79} described
a portable peephole optimizer:
\begin{quote}
Given an assembly language program and a symbolic machine description,
PO simulates pairs of adjacent instructions and, where possible,
replaces them with an equivalent single instruction.
\end{quote}
This work is particularly impressive given that automated theorem
provers were primitive when it was published.


\newcommand\firm{\footnote{\url{https://github.com/libfirm/libfirm}}}
\newcommand{\ra}[1]{\renewcommand{\arraystretch}{#1}}

\begin{table}[tbp]
\centering
\setlength{\tabcolsep}{3pt}
\begin{tabular}{@{}lllll@{}}
\toprule
                                     &        & target       & LHS          & search        \\
                                     & sound? & language     & extraction   &        method \\
\midrule
Fraser~\cite{Fraser79}               & yes    & asm          & control flow & exhaustive    \\
Davidson and Fraser\cite{Davidson84} & yes    & asm          & data flow    & exhaustive    \\
Massalin~\cite{Massalin87}           & no     & asm          & by hand      & exhaustive    \\
Denali~\cite{Joshi02}                & yes    & Alpha        & by hand      & synthesis     \\
Bansal and Aiken~\cite{Bansal06}     & yes    & x86          & control flow & exhaustive    \\
Sands$^a$                            & no     & LLVM IR      & data flow    & exhaustive    \\
STOKE~\cite{Schkufza13}              & yes    & x86-64       & by hand      & randomized    \\
Optgen~\cite{Buchwald15}             & yes    & FIRM IR      & data flow    & exhaustive    \\
Souper~\cite{Sasnauskas17}           & yes    & LLVM IR      & control + data & synthesis     \\
\bottomrule
\end{tabular}}
\vskip 0.5em
\footnotesize{$^a$~{\url{https://www.youtube.com/watch?v=8TLbP_XTJWQ}}
\caption{Some superoptimizers}
\label{tab:supers}
\end{table}

Table~\ref{tab:supers} summarizes the design points of some prior work.
A superoptimizer is sound if (leaving aside implementation defects) it
only derives correct optimizations.
Unsound tools---that sometimes derive incorrect optimizations---exist
because it may be much faster to simply test the original and
optimized code against each other, rather than invoking an automated
theorem prover.
Bansal and Aiken's tool~\cite{Bansal06} used a hybrid strategy: it
rapidly ruled out obviously incorrect optimizations via testing but
fell back on a solver to achieve soundness.
A blog post by
Sharp\footnote{\url{https://jamey.thesharps.us/2017/06/19/search-based-compiler-code-generation/}}
contains additional discussion of search-based code generation.

Historically, most work on peephole optimization has been at the level
of machine instructions, with the goal of cleaning up routine
inefficiencies emitted by relatively simple code generators.
On the other hand, in 1982 Tanenbaum~\cite{Tanenbaum82} said that
\begin{quote}
... it is desirable to do as much optimization as possible on the
intermediate code, because that optimizer can be written once and for
all and used without change as a filter for subsequent front ends and
back ends.
\end{quote}
In practice it looks like IR and backend superoptimizers are both
useful and desirable; we'll have more to say about this later.

The left-hand side (LHS) of a compiler optimization is the code that
will be optimized.
The ``LHS extraction'' column of Table~\ref{fig:big} answers the
question: How does the superoptimizer extract the program fragments
that it will attempt to optimize?
The easiest answer is ``this is done by hand,'' which is suitable for
superoptimizers like STOKE and Massalin's that are primarily aimed at
aiding assembly language programmers and library developers.
For a superoptimizer that runs as part of a compiler, extraction may
be via control flow (adjacent instructions are optimized) or data
flow (dependent instructions are optimized).
The latter is likely to be more effective, since it is insensitive to
accidents of instruction layout, but the former strategy may be useful
in particularly simple or just-in-time optimizers.
The search method is how a superoptimizer finds the cheapest
right-hand side that refines the given LHS\@.


%% file: declarative.tex
\section{Case Study Part 1: Declarative Peephole Optimizations for LLVM}
\label{sec:declarative}

As a running example, let's look at taking \texttt{((x << 31) >> 31) +
1}, an inefficient idiom for isolating and flipping the low bit of a
signed 32-bit integer, and rewriting it as \texttt{\textasciitilde
x \& 1}.
In LLVM IR, the LHS of this optimization is:

{\small\begin{verbatim}
%2 = shl i32 %0, 31
%3 = ashr i32 %2, 31
%4 = add nsw i32 %3, 1
\end{verbatim}}

\noindent
and the RHS is:

{\small\begin{verbatim}
%2 = xor i32 %0, -1
%3 = and i32 %2, 1
\end{verbatim}}

\noindent
In InstCombine, this optimization, which depends on executing in a
context where the current instruction is already known to be adding
one to \texttt{Op0}, is:

{\small\begin{verbatim}
const APInt *C3;
if (match(Op0, m_AShr(m_Shl(m_Value(X), m_APInt(C2)), m_APInt(C3))) &&
    C2 == C3 && *C2 == Ty->getScalarSizeInBits() - 1) {
  Value *NotX = Builder.CreateNot(X);
  return BinaryOperator::CreateAnd(NotX, ConstantInt::get(Ty, 1));
}
\end{verbatim}}

The transformation has two parts: a condition that looks for
optimizable code (e.g., the \texttt{m\_AShr} function pattern-matches
an arithmetic right-shift) and a body that creates new instructions to
replace the old ones.

Problems that stem from writing peephole optimizations in C++ include:
\begin{itemize}
\item
Since shared computations are factored out of optimizations for
 efficiency, there is substantial entanglement across optimizations.
\item
Automated verification of transformations written in imperative code
is challenging.
Verification is desirable because InstCombine transformations are very
easy to get wrong.
\item
Optimizations contain individual, customized reasoning about
profitability, making it difficult to maintain consistency or to
experiment with alternate profitability heuristics.
\item
There exist groups of optimizations that are individually
unprofitable, that become profitable when they can be performed
together.
LLVM's recognition of these situations, and its ability to act upon
them, is ad hoc at best.
\item
Many optimizations conservatively and unnecessarily drop instruction
attributes about undefined behavior; a formal verification tool can
easily tell developers when these flags can be preserved or added.
\item
Since its application strategy is entwined with its transformations,
speeding up InstCombine is not easy.
\item
Termination problems can occur when optimizations undo each
other~\cite{Menendez16}.
\end{itemize}


The declarative optimization language needs to be easy for LLVM
developers to read and write, and it must allow a collection of tools
to be written around it.
Alive~\cite{Lopes15} is an example of what such a language could look
like; in it, the optimization above is:

{\small\begin{verbatim}
Pre: C == width(%in) - 1
%1 = shl %in, C
%2 = ashr %1, C
%out = add %2, 1
  =>
%3 = and %in, 1
%out = xor %3, 1
\end{verbatim}}

Although Alive was primarily designed for expressiveness and formal
verification, we also conducted an experiment in automatically
converting Alive patterns into C++ code performing the specified
optimizations.
This worked, but the generated code simply iterated over the rules;
there is much room for improvement in terms of factoring out common
code, detecting collections of optimizations that need to be applied
together, etc.
A suitable framework for this might be a multiple subtree matching
algorithm~\cite{Flouri10,Sekar92}, similar to well-known automata-based
substring matching techniques; it automatically avoids duplicated work
when similar subtrees are being searched for.
The automaton will be generated from the list of declarative
optimizations at compiler-compile-time.

Another problem that can be solved naturally within this framework is
recognizing optimizations that are resistant to the greedy application
strategy because they are unprofitable individually but profitable
when performed together.
We'll use this function to illustrate the issue:

{\small\begin{verbatim}
unsigned foo(unsigned a, unsigned b) {
  unsigned na = -a;
  unsigned nb = -b;
  unsigned c = na - nb;
  unsigned d = na + nb;
  return c ^ d;
}
\end{verbatim}}

\noindent
This trivially compiles into five arithmetic operations:

{\small\begin{verbatim}
define i32 @foo(i32, i32) {
  %3 = sub i32 0, %0
  %4 = sub i32 0, %1
  %5 = sub i32 %3, %4
  %6 = add i32 %3, %4
  %7 = xor i32 %5, %6
  ret i32 %7
}
\end{verbatim}}

\noindent
At this point, there are two InstCombine transformations that apply.
First, ${-a} - {-b}$ can be rewritten as $b - a$, saving two operations.
Second, $-a + -b$ can be rewritten as $-(a - b)$, saving one operation.
The resulting function will contain four arithmetic instructions:

{\small\begin{verbatim}
define i32 @foo(i32, i32) {
  %3 = sub i32 %1, %0
  %4 = add i32 %0, %1
  %5 = sub i32 0, %4
  %6 = xor i32 %3, %5
  ret i32 %6
}
\end{verbatim}}

The problem is that, when considered individually, neither of these
transformations looks profitable, because they both add new
instructions while not obviously removing old ones: each of them
eliminates uses of $-a$ and $-b$ but those values are used by the
other code.
In cases like this, the author of an InstCombine transformation has
two choices.
First, avoid performing a rewrite when instructions on the LHS have
external uses.
Second, go ahead and perform the rewrite, on the optimistic assumption
that some other optimization will eliminate the extra uses.
In this case, an LLVM contributor made the optimistic assumption and
the gamble pays off, allowing InstCombine to produce the
four-operation version of this function.
However, it is easy to construct code where the optimism is unfounded.
This C code (where \texttt{x} and \texttt{y} are global unsigned
ints):

{\small\begin{verbatim}
unsigned bar(unsigned a, unsigned b) {
  unsigned na = -a;
  x = na;
  unsigned nb = -b;
  y = nb;
  unsigned d = na + nb;
  return d;
}
\end{verbatim}}

\noindent
compiles to:

{\small\begin{verbatim}
define i32 @bar(i32, i32) {
  %3 = sub i32 0, %0
  store i32 %3, i32* @x
  %4 = sub i32 0, %1
  store i32 %4, i32* @y
  %5 = add i32 %3, %4
  ret i32 %5
}
\end{verbatim}}

\noindent
and then InstCombine optimistically performs a transformation,
increasing the number of arithmetic operations from three to four:

{\small\begin{verbatim}
define i32 @bar(i32, i32) {
  %3 = sub i32 0, %0
  store i32 %3, i32* @x
  %4 = sub i32 0, %1
  store i32 %4, i32* @y
  %5 = add i32 %0, %1
  %6 = sub i32 0, %5
  ret i32 %6
}
\end{verbatim}}

In contrast, we would like a peephole optimization pass to have the
property that it never gratuitously adds a useless instruction.
A solution is to speculatively perform all possible transformations on
a function, tracking which instructions' use counts go to zero, and
then committing only to optimizations that, performed together, give a
global win.%
\footnote{https://twitter.com/johnregehr/status/942094482828181504}
This strategy, however, only works within InstCombine; a more
comprehensive approach such as equality saturation~\cite{Tate11} will be
needed to solve coordination problems across passes.
In conclusion:
\begin{thesis}
When possible, compilers should specify rules, machine
characteristics, and other regular information in declarative formats,
in order to facilitate rapid updating, independent checking, and
efficient translation to code that executes when the compiler runs.
\end{thesis}


%% file: souper.tex
\section{Case Study Part 2: Superoptimizing LLVM}
\label{sec:souper}


Once the declarative language and its associated tooling is in place,
we are free to strengthen the optimizer further by writing more rules
by hand, without fear of miscompilation or of slowing down the
compiler much.
So why not stop here?
There are several reasons to prefer automated derivation of
optimizations.
First, LLVM is being targeted by new programming languages: Rust,
Julia, and others.
Not only is LLVM not particularly tuned to optimize patterns that are
commonly produced by frontends for these languages, but also
higher-level languages tend to lean more heavily on the optimizer than
do C and C++.
Second, sometimes we need to adjust the semantics of IR constructs.
As of fall 2018 there are undefined-behavior-related issues, that are
still being resolved~\cite{Lee17}, that are going to eventually
invalidate some optimizations currently in LLVM while enabling new, as
yet unimplemented, transformations in InstCombine.
Third, useful optimizations for C and C++ are still missing.
In all of these cases, we can probably find better ways to use
compiler developers' time than reading optimized IR and trying to find
missed optimizations in it.

Souper~\cite{Sasnauskas17} is a superoptimizer for LLVM IR that
derives optimizations similar to the ones in InstCombine.
Souper works by:
\begin{enumerate}
\item
Choosing a ``root'' SSA value that it will attempt to compute more
cheaply.
\item
Recursively following backwards edges in the SSA graph, extracting
instructions until it is blocked by a function entry, a loop, or an
unsupported instruction (floating point, function call, load from
memory, and a few others).
Souper also tracks information learned from diverging and converging
control flow edges using, respectively, path conditions and ``block
path conditions.''
\item
Attempting to find a cheaper way to compute the value using
counterexample-guided inductive synthesis (CEGIS)~\cite{Gulwani11}.
\end{enumerate}
Eventually, Souper will attempt to optimize every integer-typed SSA
value.
Since CEGIS is not fast, this can be very time consuming, particularly
if the solver is allowed to run for a while before timing out and if
we attempt to synthesize relatively large RHSs.
However, synthesis results are cached and can subsequently be applied
relatively quickly: in a preliminary experiment we saw Souper
increasing compile time by about 10\% in the warm-cache case, compared
to a regular \texttt{-O3} compile.
After compiling LLVM itself (3.5 MSLOC of C++) Souper's cache occupies
362\,MB of RAM; when dumped to disk, the file is 149\,MB.
This is acceptable for a research prototype but it isn't the solution
that we want to deploy.

Returning to the running example, its LHS in Souper IR is very similar
to the LLVM version:

{\small\begin{verbatim}
%in:i32 = var
%1 = shl %in, 31
%2 = ashr %1, 31
%out = addnsw %2, 1
infer %out
\end{verbatim}}

\noindent
Souper's synthesis produces this RHS:

{\small\begin{verbatim}
%4:i32 = xor 1:i32, %in
%5:i32 = and 1:i32, %4
result %5
\end{verbatim}}

This optimization, like every optimization discovered by Souper, is
completely specific: it only applies to one pattern of instructions,
one choice of values for constants, one width of operators, etc.
It would be preferable to exploit---as peephole optimizers written by
humans do---the fact that optimizations are almost always more broadly
applicable.
Generalization can be done along multiple axes, here we'll look at
relaxing constraints on bitwidths and choice of constants.
Starting with the optimization derived by Souper above, we can
translate it into Alive like this:

{\small\begin{verbatim}
%1 = shl i32 %in, 31
%2 = ashr %1, 31
%out = add nsw %2, 1
  =>
%4 = xor 1, %in
%out = and 1, %4
\end{verbatim}}

\noindent
The \texttt{i32} bitwidth constraint in the first line suffices to
constrain the entire optimization to the 32-bit case.
This optimization can be trivially generalized by removing the
bitwidth constraint and by replacing each constant on the LHS with a
symbolic constant:

{\small\begin{verbatim}
%1 = shl %in, C1
%2 = ashr %1, C2
%out = add nsw %2, C3
  =>
%4 = xor 1, %in
%out = and 1, %4
\end{verbatim}}

\noindent
The problem with this more generic optimization is that it doesn't
work for all choices of constants: it requires a precondition check
before it can fire safely
Menendez and Nagarakatte~\cite{Menendez17} showed that it is possible
to automatically derive weakest preconditions for Alive optimizations.
For this example, their tool, Alive-Infer, comes up with:

{\small\begin{verbatim}
(((~C2 | ~C1) == -width(%out)) && (C3 == 1))
\end{verbatim}}

\noindent
Although this looks a little funny, and it isn't minimal in terms of
operations performed, it is indeed a weakest precondition for the
optimization to fire.
%

There are other dimensions along which an optimization can be generalized.
It is sometimes the case that when LLVM's undefined behavior
qualifiers are present on the LHS, they can be preserved on the RHS\@.
The declarative optimization language should (as Alive does)
specifically support automatically tagging RHSs with as many of these
flags as soundness allows.
Another form of generality is found in optimizations that
want to perform similar transformations across a range of different
instructions.
For example, in InstCombine it is common to see an optimization
 involving comparison instructions that is parameterized by the
 comparison type.
This could be supported in the optimization language using a
regular-expression-like mechanism, or alternatively we could simply
require each pattern to be specified separately.

Souper appears to be good at a few things besides finding peephole
optimizations.
First, it can robustly recognize idioms such as rotate and Hamming
weight computation that span multiple instructions.
In contrast, compiler idiom recognizers tend to be fragile.
For example, whereas optimizing C and C++ compilers have often been
good at turning the obvious code \texttt{x << r | x >> (32 - r)} into
a 32-bit rotate-left instruction, they generally failed to recognize
the slightly more complicated code that must be used to avoid
undefined behavior in the rotate-by-zero case.%
\footnote{\url{https://blog.regehr.org/archives/1054} and
\url{https://blog.regehr.org/archives/1063}}
Idiom recognition is useful when a source or intermediate language
does not support direct expression of the idiom.
Souper is good at this because the SAT solver is effective at
deobfuscating whatever code ends up being written by humans.
Second, Souper is good at finding dead code; we've seen Souper reduce
the size of a Clang executable by about 3\,MB (4.4\%).

In summary, although many of the pieces are in place, a
superoptimizer-generated InstCombine replacement for LLVM will require
both research and engineering work.
On the other hand, a more limited goal---a research prototype
replacing InstCombine with the Souper LLVM pass, omitting the
generalization, fast matching, and declarative optimization
language---is within easy reach.

Souper itself has many areas that need improvement.
It should support memory instead of living only in the SSA world.
It should support vector instructions, floating point instructions,
loops, and it should look across function boundaries.
Finally, Souper is sometimes limited by the capabilities of the SMT
solver, particularly when divisions or floating point operations are
involved (an extremely preliminary FP-aware Souper prototype
exists).
In summary:
\begin{thesis}
Declarative specifications for optimizations (and other elements of
 compilers) are an appropriate interface between offline search-based
 tools and the compiler itself.
\end{thesis}


%% file: moresuperoptimizers.tex
\section{Synthesizing More of Compilers Automatically}

The scope of an IR$\rightarrow$IR superoptimizer is limited: it cannot
exploit target-specific optimizations such as fun addressing
modes, nor does it have access to PL-level information, such as types,
that is often necessary to perform higher-level optimizations.
How can we get solver-driven optimization at higher and lower levels
than IR?
The answer is clear: more superoptimizers.

The case for target-aware superoptimization has already been made in
the context of offline tools that come up with suggestions for
compiler implementors.
For example, Granlund and Kenner~\cite{Granlund92} report that with
the GNU superoptimizer:
\begin{quote}
A number of surprising results were obtained, many of which were
unknown to the architects of the RS/6000 processor.
\end{quote}

Another technique with a long history is facilitating the
development of compiler backends by writing incomplete machine
descriptions in a declarative language, instead of embedding this
information in imperative code.
LLVM and GCC both do this.%
\footnote{\url{https://llvm.org/docs/TableGen/index.html}\\
\url{https://gcc.gnu.org/onlinedocs/gccint/Machine-Desc.html\#Machine-Desc}}
These descriptions serve as a partial solution to the never-ending
tasks of creating a backend for each CPU or GPU target that must be
supported, and adjusting and retuning these backends with new pipeline
models, auxiliary instructions, etc.\ as new micro-architectures are
released.
A logical step forward would be to augment these structural
instruction descriptions with formal semantics of the instructions
being described, as a step towards solver-based generation of major
backend components such as instruction selectors.
This idea has been explored in a number of research
projects~\cite{Dias10,Jangda17,Buchwald18}.
A goal for future work is to automate the construction of as much of a
compiler backend as possible.
In situations where compile time is not a significant constraint, a
solver can be used in an online fashion, allowing optimality
guarantees to be made.
However, in the common case the backend will need to be generated offline,
even if this results in missed code generation opportunities.

Some compilers have a language-specific IR, such as Rust's
MIR\footnote{\url{https://blog.rust-lang.org/2016/04/19/MIR.html}} and
Swift's
SIL,\footnote{\url{https://github.com/apple/swift/blob/master/docs/SIL.rst}}
that is intended to facilitate optimizations that require more
source-level information.
These should be targeted by superoptimizers.
In other words:

\begin{thesis}
Every IR in a compiler, and every translation between representations,
is an opportunity for superoptimization.
\end{thesis}

Beyond front-end, middle-end, and backend optimizers, more parts of
compilers can be generated with help from solvers.
For example, optimizers rely heavily on static analyses, which can be
viewed as ad hoc, domain-specific automated theorem provers.
For instance, LLVM has ``basic alias
analysis,''\footnote{\url{https://github.com/llvm-mirror/llvm/blob/release_60/lib/Analysis/BasicAliasAnalysis.cpp}}
that uses relatively simple rules to reason about pointers, and also
value tracking, which attempts to prove that individual bits have
fixed
values.\footnote{\url{https://github.com/llvm-mirror/llvm/blob/release_60/lib/Analysis/ValueTracking.cpp}}
Each of these solvers takes a collection of rules that express locally obvious
properties of code (e.g., ``a pointer to a freshly allocated memory
block cannot alias any existing block'') and pushes them around the
program with the goal of, for example, proving that a store instruction
does not write to the same memory region as another store.

The alias analysis rules in LLVM are embedded in the C++ code
implementing the pass.
Alternatively, these rules could be written mathematically.
This will make it easier to prove that each rule is consistent with
the semantics of LLVM IR, while still being possible to automatically
generate C++ code that implements a given analysis that propagates
those rules.

Static analysis rules themselves can also be generated
automatically~\cite{scherpelz07}.
Given the semantics of the IR and the semantics of the result of the
analysis, one can use an SMT solver to automatically generate the most
precise transfer functions that implement the analysis.

%% file: data.tex
\section{Data-Driven, not Intuitive, Compiler Design and Implementation}
\label{sec:datadriven}


As large, complex artifacts, compilers have significant inertia: they
are better at compiling yesterday's programs for yesterday's
architectures than they are for solving today's problems.
As we heard a GCC developer say: ``it's like piloting a
supertanker---we can steer, but it takes a really long time to change
direction.''
The inertia is rooted in thousands of little design decisions that
ossify once they get implemented.
For example, consider a relatively simple compiler component: a
conservative static analysis of the values that an integer-typed
variable might take.
Before implementing this analysis we'll need to decide:
\begin{itemize}
\item
Should it be lazy or eager?
\item
Should its results be cached and, if so, at what granularity should
the cache be invalidated?
\item
Should integer ranges be allowed to wrap?
\item
Should the analysis be relational?
\item
Should it be flow, context, or path sensitive, and if so to what degree?
\item
Should it attempt to retain precision when confronted with bitwise
operators, and if so how hard should it try?
\item
Etc.
\end{itemize}
Answering these questions requires a detailed understanding of how the
analysis results will be used, the characteristics of the programs
being analyzed, and the available resources at compile time.
Some of the answers will be based on experimental data but others will
come from a developer's intuition.
Eventually the developer will move on to other jobs and at the same
time there will be drift in the character of the programs being
compiled, the target architectures, and the rest of the compiler's
structure and capabilities.
As a specific example, many design and implementation decisions in
LLVM boil down to a guess that Chris Lattner made ca.~2007; some of
these have aged well while others have not.

A compiler solves both \emph{hard} and \emph{soft} problems.
A hard problem is one where a mistake threatens the compiler's
correctness.
The implementation of almost any optimization or static analysis
involves solving hard problems.
A soft problem is one where a mistake potentially affects the resource
usage of the compiler or the compiled program, but does not threaten
correctness.
Soft problems include deciding which optimization passes to run and in
what order, which specific optimizations to perform, how hard a static
analysis should try to reach a useful conclusion, etc.
This leads to:

\begin{thesis}
A data-driven approach should be used to solve both hard and soft
problems in compilers.
\end{thesis}

Sources of data include the programs being compiled, execution
characteristics of these programs on targets of interest, and
execution characteristics of the compiler itself.
Hard problems are best attacked by discrete methods such as
automated theorem provers.
Soft problems are best attacked by continuous methods such as
combinatorial optimization and machine learning.

A part of a compiler is data-driven if, given enough data, it can
automatically and rapidly (perhaps within a few hours or days) be
re-tuned to fit new circumstances.
Most parts of most current compilers do not meet this criterion.

Although hard compiler problems can be solved using data---the
scenario in Section~\ref{sec:souper} where Souper is used to
synthesize missing optimizations is an example---soft problems are the
more obvious targets.
For example, effective compilation of modern C++ requires good
heuristics about when to inline a function call.
These heuristics are relatively difficult to get right and are not
often considered to be completely satisfying.
Moreover, when LLVM gets targeted by a new language, such as Rust or
Julia, or is itself retargeted to a new platform, such as an MSP430
with extreme code size constraints, it is not trivial to retune the
inliner for the new situation.
A data-driven approach to inlining would, in contrast, run many
experiments in order to derive an inlining strategy that maximizes an
objective function such as ``make Julia code fast'' or ``reduce code
size on an MSP430.''
These experiments will require a large amount of input code and may
be computationally expensive, but they can operate on a time scale
that is a small fraction of that required for a significant
engineering effort.
Considerable research exists on the data-driven approach to solving
soft compiler problems, including finding a good phase
order~\cite{milepost}, automatic heuristic generation~\cite{Simon13},
and algorithm autotuning~\cite{opentuner}.


%% file: semantics.tex
\section{We Need First-Class Formal Semantics}

Many important program representations such as C, C++, x86, and the IRs
for LLVM, GCC, and Visual C++, either lack a formal semantics or else
lack a first-class formal semantics.
This situation impedes the development of formal-methods-based tools
because the authors of these tools are forced to formalize a program
representation before implementing the tool itself.
The resulting semantics tend to be one-offs: they are embedded in a
tool, tailored for a specific purpose, and usually cannot be easily
reused.
The process of formalization is itself (ironically) error-prone and,
worse, it often exposes latent ambiguities in real-world systems that
can be hard to resolve definitively.
For example, our work on undefined behavior in LLVM IR~\cite{Lee17}
ran into this kind of issue, and several years later we're still
working with the LLVM community to get them resolved.

A first-class formal semantics is one that:
\begin{itemize}
\item
Stands on its own, independent of any particular tool or use case.
\item
Is widely agreed to be authoritative: any defect in the semantics is
fixed with high priority and deviations from the semantics are incorrect
by definition.
\item
Is at the root of an ecosystem of mechanically derived artifacts:
documentation, simulators, parsers, static analyzers, compilers and
decompilers, etc.
\item
Has clients both above and below in the tool stack. For example, an
x86 semantics is used from above when proving a compiler backend
correct, and is used from below when proving that a chip faithfully
implements the architecture.
\item
Has people whose job descriptions include taking care of it: answering
questions, keeping it up to date, refactoring it to make it easier to
use, writing special-purpose tools to look for missing cases,
vacuous elements, and other defects.
\end{itemize}
These criteria are met, or nearly met, by a semantics for version 8.3
of the Arm architecture.%
\footnote{\url{https://alastairreid.github.io/arm-v8_3/}}
In some cases, the technology necessary to create a usable first-class
formal semantics may not yet exist, and in other cases the cost of
creating the semantics will not be small.
For example, C++ probably has both of these problems.

\begin{thesis}
Durable interfaces, such as ISAs, IRs, and programming languages,
should be accompanied by first-class formal semantics.
\end{thesis}

An important use case for formal semantics is translation validation:
a proof that a compiler's output refines its input.
The huge advantage of translation validation over invasive proofs of
compiler correctness is that translation validation does not ask
compiler developers to discharge proof obligations.
Therefore, compiler users with high confidence requirements
(e.g.\ those developing avionics software) and compilers users with
lower confidence requirements can share the same compiler
infrastructure, and only the high-confidence users need to pay the
increased tool development and CPU time costs of translation
validation.

In principle, a separate translation validation tool is unnecessary
when a compiler is made of verified pieces.
In practice, it is going to be a long time before all of the pieces
are verified: there are many ways to go wrong while composing verified
transformations (particularly when the implementation language is
unsafe), and also a redundant end-to-end check provides some defense
in depth against otherwise-undetected defects.

Translation validation isn't only for high-confidence use cases, it is
also a natural fit for a compiler testing campaign.
The process is easy: compile a piece of code (either extracted from applications
or generated automatically) and then try to prove that the optimized code refines
the original code.
We found several LLVM bugs by doing this for small, automatically
generated functions.%
\footnote{\url{https://blog.regehr.org/archives/1510}}
Compiler bugs can be very difficult to flush out with testing; our
experience is that translation validation tools can be an effective
way to find these bugs, though of course the problem of triggering the
buggy compiler optimizations remains.


%% file: further.tex
\section{Pushing Solver-Based Compiler Implementation Further}
\label{sec:futher}

\paragraph{Integrated analyses.}
Typically, a static analysis is based on an abstract domain such as
integer ranges, polyhedra, or points-to sets.
The analysis proceeds by applying a collection of abstract transfer
functions, each describing the abstract effect of some concrete part
of the program, until a fixed point is reached.
In realistic situations, analysis precision is dropped on the floor in
cases where two abstract domains could learn from each other, but they
have not been taught to do so.
Teaching static analyses to learn from each other is a highly
demanding task and typically it is only done in very limited cases
such as sparse conditional constant propagation: the most precise
combination of constant propagation and dead code
elimination~\cite{Wegman91}.
On the other hand, solvers are very good at recognizing special cases
such as those that allow extra precision to be gleaned from
interacting dataflow facts; the resulting integrated analyses will
lead to compilers that optimize more robustly and have fewer
phase-ordering problems.

\paragraph{Reducing myopia in transformations.}
Optimizers tend to be very effective when every optimization step is
relatively simple and is clearly a good idea.
However, not all optimizations have this character: sometimes the
local gradient points in the wrong direction and a substantial piece
of code must be recognized as a whole, so that it can be replaced with
something more efficient.
Optimizers therefore contain custom code recognizing common
implementations for idioms---such as integer overflow checks, rotates,
byte shuffles, and Hamming weight computations---that are efficient at
the CPU level but cannot be conveniently expressed in most programming
languages.
This detection, however, is often not thorough: a few patterns that
have been seen in practice (or in benchmarks) are detected, but an
unknown number of less important codes fail to get the performance
benefits because they express the inefficient operations differently.
In contrast, solvers are not easily fooled by incidental changes in
the structure of a computation and they can be used as the basis for
more robust idiom detectors.
For example, we have had good luck recognizing Hamming weight
computations using Souper (though only when they are loop-free).
Inefficient sorting algorithms and numerical algorithms seem
like good targets for future work in this direction.

\paragraph{Data structure synthesis.}
Codes in high-level languages often make heavy use of APIs such as
container classes or tensors.
However, the particular choices made by library implementors are not
always suitable for performance-critical use cases.
One workaround is to use customized libraries; for example, LLVM has
``small'' versions of the C++ set, vector, and other container classes
that avoid heap allocation as long as the container does not grow
beyond a small, predetermined number of elements.
Alternatively, given an API and a collection of profile-like data
about how the API is used, we could try to synthesize a more efficient
data structure along with its collection of accessors and mutators.
Progress in this direction exists~\cite{Loncaric18} but it's fair to
say that it will be a while before developers can routinely replace
elements of standard libraries with superior, synthesized
alternatives.

\paragraph{User-defined optimizations.}
One of Robison's points~\cite{Robison01} is that there are many
programs that would benefit from domain-specific compiler
optimizations, but that the economics of compiler development are
often unfavorable.
While some parts of this problem will be ameliorated by the
data-driven approach---given a sufficient body of code, a compiler
specifically tuned for an application niche can be automatically
constructed---and other parts have been solved by open source
compilers, the basic problem of teaching a compiler new tricks
remains.
We believe that domain-specific optimization languages and their
associated proof machinery, along the lines of the example in
Section~\ref{sec:declarative}, are a promising approach.


%% file: ir.tex
\section{Next-Generation IRs}

\paragraph{The IR instruction set.}
Adding an instruction to a compiler IR can be a bit painful: every
backend needs to accommodate the new instruction and also various
analyses and transformations in the middle-end need to be taught what
the new instruction means, or else there will be code quality
regressions.
In contrast, the parts of a compiler that are derived using a solver
will not require this kind of manual adjustment, since their ideas
about the meaning of the new instruction come from its formal
semantics, not from code written by people to cope, individually, with
each situation in which the new instruction's meaning is relevant.
At the same time that it should be easier to add instructions to a
solver-based compiler, the necessity to do so may decrease due to the
ease (discussed in Section~\ref{sec:futher}) with which solvers can
recognize sophisticated idioms hiding in collections of simple
instructions.

\paragraph{Reducing pointer chasing.}
Since the in-memory version of a compiler IR is typically
pointer-intensive, traversing IR requires a lot of indirections.
Informal measurements show that on a modern core with exclusive use of
a 25\,MB cache, an optimizing C++ compile using GCC or LLVM still
spends 30--35\% of its runtime stalled on memory operations.
The pointer-heavy representation follows from the desire to easily
edit the IR on the fly using hand-written passes.

Future IRs that are less-often manipulated in ad hoc fashion can be be
designed to be cache friendly and also suitable for processing using
GPUs and GPU-like many-core processors.
As an example of this kind of work, Vollmer et al.~\cite{Vollmer17}
report that ``For traversals touching the whole tree, such as maps and
folds, packed data allows speedups of over 2× compared to a
highly-optimized pointer-based baseline.''

\paragraph{Abstraction.}

Over the years, IRs have tended to become more abstract and amenable
to mathematical analysis, retaining fewer and fewer incidental aspects
specified by the original computation.
For example:
\begin{itemize}
\item register transfer language (RTL)~\cite{Davidson84} allowed peephole
  optimizers to be machine-independent
\item static single assignment (SSA)~\cite{Cytron91} eliminates some
  kinds of incidental mutation, making dependency information much
  easier to discover
\item value state dependence graph (VSDG)~\cite{Lawrence07} and
  related IRs such as sea of nodes~\cite{Click96} don't store
  instructions in lists, but rather make all dependencies explicit
\item in LLVM IR, integer operations can take any bitwidth, as opposed
  to being restricted to widths supported by machine instructions.
\end{itemize}
The general trend is that a more abstract IR can make desirable
optimizations more convenient to perform, but then more work is
required to convert the code into a readily executable form.
A more abstract IR may not always be a win in terms of code quality:
we heard an anecdote where a VSDG-like IR turned out to be a
showstopper because it would lose instruction ordering information
inserted by expert programmers, and then the compiler backend was
incapable of independently rediscovering the desirable ordering.
Intuitively, we can expect a solver-oriented compiler to be a better
match for a more abstract IR, and we can hope that its backends will
be smart enough to cope with the additional abstraction.


%% file: conclusions.tex
\section{Conclusions}

This paper outlines an agenda for making compilers better by
incrementally removing hand-written parts and replacing them with
components derived using automated theorem provers, formal semantics,
and data-driven tools.
Although making this happen will require solving many interesting
research problems, we have focused on the engineering advantages of
the proposed approach, which we believe are clear and significant.


%% file: paper.bbl
\begin{thebibliography}{10}

\bibitem{opentuner}
Jason Ansel, Shoaib Kamil, Kalyan Veeramachaneni, Jonathan Ragan-Kelley,
  Jeffrey Bosboom, Una-May O'Reilly, and Saman Amarasinghe.
\newblock Opentuner: An extensible framework for program autotuning.
\newblock In {\em PACT}, 2014.
\newblock
  \url{http://groups.csail.mit.edu/commit/papers/2014/ansel-pact14-opentuner.pdf}.

\bibitem{Bansal06}
Sorav Bansal and Alex Aiken.
\newblock Automatic generation of peephole superoptimizers.
\newblock In {\em ASPLOS}, 2006.
\newblock
  \url{http://theory.stanford.edu/~aiken/publications/papers/asplos06.pdf}.

\bibitem{Buchwald15}
Sebastian Buchwald.
\newblock Optgen: A generator for local optimizations.
\newblock In {\em CC}, 2015.
\newblock \url{https://pp.ipd.kit.edu/uploads/publikationen/buchwald15cc.pdf}.

\bibitem{Buchwald18}
Sebastian Buchwald, Andreas Fried, and Sebastian Hack.
\newblock Synthesizing an instruction selection rule library from semantic
  specifications.
\newblock In {\em CGO}, 2018.
\newblock \url{https://pp.ipd.kit.edu/uploads/publikationen/buchwald18cgo.pdf}.

\bibitem{Click96}
Clifford Click, Jr.
\newblock {\em Combining Analyses, Combining Optimizations}.
\newblock PhD thesis, Rice University, 1995.

\bibitem{Cytron91}
Ron Cytron, Jeanne Ferrante, Barry~K. Rosen, Mark~N. Wegman, and F.~Kenneth
  Zadeck.
\newblock Efficiently computing static single assignment form and the control
  dependence graph.
\newblock {\em ACM Trans. Program. Lang. Syst.}, 13(4), October 1991.

\bibitem{Davidson84}
Jack~W. Davidson and Christopher~W. Fraser.
\newblock Code selection through object code optimization.
\newblock {\em ACM Trans. Program. Lang. Syst.}, 6(4), October 1984.
\newblock
  \url{https://people.well.com/user/cwf/pro/Davidson%20and%20Fraser.%20Code%20selection%20through%20object%20code%20optimization.pdf}.

\bibitem{Dias10}
Jo\~{a}o Dias and Norman Ramsey.
\newblock Automatically generating instruction selectors using declarative
  machine descriptions.
\newblock In {\em POPL}, 2010.
\newblock \url{https://www.cs.tufts.edu/~nr/pubs/gentileset.pdf}.

\bibitem{Eide08}
Eric Eide and John Regehr.
\newblock Volatiles are miscompiled, and what to do about it.
\newblock In {\em EMSOFT}, 2008.
\newblock \url{http://www.cs.utah.edu/~regehr/papers/emsoft08-preprint.pdf}.

\bibitem{Flouri10}
Tomáš Flouri, Jan Janoušek, and Bořivoj Melichar.
\newblock Subtree matching by pushdown automata.
\newblock {\em Computer Science and Information Systems}, 48(14):331--357,
  2010.

\bibitem{Fraser79}
Christopher~W. Fraser.
\newblock A compact, machine-independent peephole optimizer.
\newblock In {\em POPL}, 1979.
\newblock
  \url{https://people.well.com/user/cwf/pro/Fraser.%20A%20compact,%20machine-independent%20peephole%20optimizer.pdf}.

\bibitem{milepost}
Grigori Fursin, Yuriy Kashnikov, Abdul~Wahid Memon, Zbigniew Chamski, Olivier
  Temam, Mircea Namolaru, Elad Yom-Tov, Bilha Mendelson, Ayal Zaks, Eric
  Courtois, Francois Bodin, Phil Barnard, Elton Ashton, Edwin Bonilla, John
  Thomson, Christopher K.~I. Williams, and Michael O'Boyle.
\newblock Milepost {GCC}: Machine learning enabled self-tuning compiler.
\newblock {\em International Journal of Parallel Programming}, 39(3):296--327,
  Jun 2011.
\newblock \url{http://homepages.inf.ed.ac.uk/ckiw/postscript/ijpp.pdf}.

\bibitem{Granlund92}
Torbj\"{o}rn Granlund and Richard Kenner.
\newblock Eliminating branches using a superoptimizer and the {GNU C} compiler.
\newblock In {\em PLDI}, 1992.

\bibitem{Gulwani11}
Sumit Gulwani, Susmit Jha, Ashish Tiwari, and Ramarathnam Venkatesan.
\newblock Synthesis of loop-free programs.
\newblock In {\em PLDI}, 2011.
\newblock
  \url{https://www.microsoft.com/en-us/research/wp-content/uploads/2016/12/pldi11-loopfree-synthesis.pdf}.

\bibitem{Jangda17}
Abhinav Jangda and Greta Yorsh.
\newblock Unbounded superoptimization.
\newblock In {\em Onward}, 2017.
\newblock \url{http://www.eecs.qmul.ac.uk/~gretay/papers/onward2017.pdf}.

\bibitem{Joshi02}
Rajeev Joshi, Greg Nelson, and Keith Randall.
\newblock Denali: a goal-directed superoptimizer.
\newblock In {\em PLDI}, 2002.

\bibitem{Lawrence07}
Alan~C. Lawrence.
\newblock {Optimizing compilation with the Value State Dependence Graph}.
\newblock Technical Report UCAM-CL-TR-705, University of Cambridge, Computer
  Laboratory, December 2007.

\bibitem{Lee17}
Juneyoung Lee, Yoonseung Kim, Youngju Song, Chung-Kil Hur, Sanjoy Das, David
  Majnemer, John Regehr, and Nuno~P. Lopes.
\newblock Taming undefined behavior in {LLVM}.
\newblock In {\em PLDI}, 2017.
\newblock \url{http://web.ist.utl.pt/nuno.lopes/pubs/undef-pldi17.pdf}.

\bibitem{Loncaric18}
Calvin Loncaric, Michael~D. Ernst, and Emina Torlak.
\newblock Generalized data structure synthesis.
\newblock In {\em ICSE}, 2018.
\newblock
  \url{https://homes.cs.washington.edu/~mernst/pubs/generalized-synthesis-icse2018.pdf}.

\bibitem{Lopes15}
Nuno~P. Lopes, David Menendez, Santosh Nagarakatte, and John Regehr.
\newblock Provably correct peephole optimizations with {Alive}.
\newblock In {\em PLDI}, 2015.
\newblock \url{http://web.ist.utl.pt/nuno.lopes/pubs/alive-pldi15.pdf}.

\bibitem{Massalin87}
Henry Massalin.
\newblock Superoptimizer: A look at the smallest program.
\newblock In {\em ASPLOS}, 1987.

\bibitem{Menendez16}
David Menendez and Santosh Nagarakatte.
\newblock Termination-checking for {LLVM} peephole optimizations.
\newblock In {\em ICSE}, 2016.

\bibitem{Menendez17}
David Menendez and Santosh Nagarakatte.
\newblock {Alive-Infer}: Data-driven precondition inference for peephole
  optimizations in {LLVM}.
\newblock In {\em PLDI}, 2017.
\newblock
  \url{https://www.cs.rutgers.edu/~santosh.nagarakatte/papers/pldi2017-alive-infer.pdf}.

\bibitem{Robison01}
Arch~D. Robison.
\newblock Impact of economics on compiler optimization.
\newblock In {\em Java Grande}, 2001.
\newblock \url{http://doi.acm.org/10.1145/376656.376751}.

\bibitem{Sasnauskas17}
Raimondas Sasnauskas, Yang Chen, Peter Collingbourne, Jeroen Ketema, Jubi
  Taneja, and John Regehr.
\newblock Souper: {A} synthesizing superoptimizer.
\newblock {\em CoRR}, abs/1711.04422, 2017.

\bibitem{scherpelz07}
Erika~Rice Scherpelz, Sorin Lerner, and Craig Chambers.
\newblock Automatic inference of optimizer flow functions from semantic
  meanings.
\newblock In {\em PLDI}, 2007.
\newblock
  \url{https://cseweb.ucsd.edu/~lerner/papers/inferring_flow_functions.pdf}.

\bibitem{Schkufza13}
Eric Schkufza, Rahul Sharma, and Alex Aiken.
\newblock Stochastic superoptimization.
\newblock In {\em ASPLOS}, 2013.

\bibitem{Sekar92}
R.~C. Sekar, R.~Ramesh, and I.~V. Ramakrishnan.
\newblock Adaptive pattern matching.
\newblock In {\em ICALP}, 1992.

\bibitem{Simon13}
Douglas Simon, John Cavazos, Christian Wimmer, and Sameer Kulkarni.
\newblock Automatic construction of inlining heuristics using machine learning.
\newblock In {\em CGO}, 2013.

\bibitem{Tanenbaum82}
Andrew~S. Tanenbaum, Hans van Staveren, and Johan~W. Stevenson.
\newblock Using peephole optimization on intermediate code.
\newblock {\em ACM Trans. Program. Lang. Syst.}, 4(1), January 1982.

\bibitem{Tate11}
Ross Tate, Michael Stepp, Zachary Tatlock, and Sorin Lerner.
\newblock Equality saturation: a new approach to optimization.
\newblock {\em Logical Methods in Computer Science}, 7(1), 2011.

\bibitem{Vollmer17}
Michael Vollmer, Sarah Spall, Buddhika Chamith, Laith Sakka, Chaitanya
  Koparkar, Milind Kulkarni, Sam Tobin-Hochstadt, and Ryan~R. Newton.
\newblock {Compiling Tree Transforms to Operate on Packed Representations}.
\newblock In {\em Proc.\ ECOOP}, 2017.

\bibitem{Wegman91}
Mark~N. Wegman and F.~Kenneth Zadeck.
\newblock Constant propagation with conditional branches.
\newblock {\em ACM Trans. Program. Lang. Syst.}, 13(2), April 1991.

\bibitem{Yang11}
Xuejun Yang, Yang Chen, Eric Eide, and John Regehr.
\newblock Finding and understanding bugs in {C} compilers.
\newblock In {\em PLDI}, 2011.
\newblock \url{http://www.cs.utah.edu/~regehr/papers/pldi11-preprint.pdf}.

\end{thebibliography}
